\documentclass[12pt,a4paper,final]{iopart}

\usepackage{iopams}  
\usepackage{graphicx}
\usepackage{cite}
\usepackage{epstopdf}
\usepackage{booktabs}
\usepackage{multirow}
\usepackage{caption}
\usepackage[breaklinks=true,colorlinks=true,linkcolor=magenta,urlcolor=magenta,citecolor=magenta]{hyperref}
\usepackage{booktabs}
\usepackage{float}
\usepackage{xcolor}

\begin{document}

\title[]{Signatures of non-trivial band topology in LaAs/LaBi heterostructure}

\author{Payal Wadhwa}
\address{TGraMS Laboratory, Department of Physics, Indian Institute of Technology Ropar, Rupnagar-140001, Punjab, India}

\author{T. J. Dhilip Kumar}%
\address{Department of Chemistry, Indian Institute of Technology Ropar, Rupnagar, Punjab-140001, India}

\author{Alok Shukla}
\address{Department of Physics, Indian Institute of Technology Bombay, Powai-400076, Mumbai, India}

\author[cor1]{Rakesh Kumar}
\address{TGraMS Laboratory, Department of Physics, Indian Institute of Technology Ropar, Rupnagar-140001, Punjab, India}
\eads{\mailto{rakesh@iitrpr.ac.in}}

\begin{abstract}
In this article, we investigate non-trivial topological features in a heterostructure of extreme magnetoresistance (XMR) materials LaAs and LaBi using density functional theory (DFT).\ The proposed heterostructure is found to be dynamically stable and shows bulk band inversion with non-trivial Z$_{2}$ topological invariant and a Dirac cone at the surface.\ In addition, its electron and hole carrier densities ratio is also calculated to investigate the possibility to possess XMR effect.\ Electrons and holes in the heterostructure are found to be nearly compensated, thereby facilitating it to be a suitable candidate for XMR studies.\

\end{abstract}


\section{Introduction}
The discovery of non-trivial topological materials like topological insulators (TI), topological superconductors (TSC), topological crystalline insulators (TCI), and topological semimetals (TSM), etc., have drawn enormous attention theoretically as well as experimentally amongst condensed matter and material science community \cite{RevModPhys.82.3045,RevModPhys.83.1057,PhysRevLett.118.236402,agarwala2019higher,PhysRevLett.106.106802,armitage2018weyl,barik2018multiple,wadhwa2019first,PhysRevB.99.205112,PhysRevLett.121.086804,PhysRevLett.121.226401}.\ TIs possess bulk band inversion and time-reversal protected surface states, which distinguish them from the conventional band insulators \cite{RevModPhys.82.3045,RevModPhys.83.1057}.\ These remarkable features of non-trivial topological phases lead to exotic topological phenomena such as Majorana fermions \cite{PhysRevB.95.155420}, quantum magnetoresistance \cite{li2016negative}, chiral anomaly \cite{PhysRevX.5.031023,vazifeh2013electromagnetic}, etc.\ To realize the phenomena and applications of TIs, we need to have a bulk band gap and protected surface states simultaneously.\ But, both of them are interdependent, therefore the attempts of achieving one may destroy the other \cite{zhang2011band,kong2011ambipolar}.\ However, the unique properties of non-trivial topological materials pave a new path to overcome the problems.\ Although bulk band inversion directly ensures the presence of protected surface states, still the specific band structure of the surface state is controlled by the prospective environment near the surface.\ It indicates that engineering a heterostructure of suitable materials with specific bulk and surface properties may provide an additional path to explore for a new non-trivial topological material \cite{PhysRevLett.115.136801,eschbach2015realization,dey2018bulk}.\\

Recently, XMR materials like WTe$_{2}$, TaAs, LaBi, PrBi, etc. \cite{ali2014large,jiang2015signature,PhysRevB.94.041103,neupane2016observation, ghimire2016magnetotransport,PhysRevB.95.115140,yu2017magnetoresistance,wang2018topological,PhysRevB.99.245131}, are found to exhibit non-trivial topological properties.\ Previous reports manifested that in topological semi-metals, surface protection by time-reversal symmetry (TRS) suppresses the electron's backscattering in the absence of magnetic field, while its presence breaks TRS, which leads to XMR effect \cite{jiang2015signature,liang2015ultrahigh}.\ In addition, XMR effect is also explained by perfect (or nearly perfect) electron-hole compensation via semi-classical two band model \cite{PhysRevLett.113.216601,sun2016large,PhysRevB.96.235128,PhysRevB.93.235142}.\ More interestingly, rare-earth monopnictides (LnPn, Ln = rare-earth element and Pn = Bi, Sb, and As) are found to exhibit XMR effect, where some of them like LaAs, LaSb, etc., are topologically trivial, while others like LaBi, CeBi, etc., are non-trivial \cite{PhysRevB.95.115140,PhysRevB.96.235128,PhysRevB.98.220102,duan2018tunable}. It is also observed that heterostructure made up of non-trivial topological materials often leads to non-trivial topological heterostructure, and may not be very surprising.\ It motivated us to investigate for topological properties in heterostructure of topologically trivial and non-trivial XMR materials, and for that we selected LaAs and LaBi.\ Since, the origin of XMR effect in LaAs and LaBi is manifested by perfect electron-hole compensation \cite{sun2016large,PhysRevB.96.235128,PhysRevB.93.235142}, therefore we further calculated the ratio of densities of electrons and holes (\textit{n$_{e}$/n$_{h}$}) of the proposed heterostructure to investigate its possibility to show XMR effect.

\section{Computational details}
DFT as implemented in VASP is used to perform all the electronic structure calculations \cite{PhysRevB.54.11169} with projected augmented wave (PAW) formalism \cite{kresse1999ultrasoft} to include electron-ion interactions.\ Perdew-Burke-Ernzehrof (PBE) functional within Generalized Gradient Approximation (GGA) and hybrid functionals HSE06 are used to incorporate exchange-correlation interactions \cite{PhysRevLett.77.3865,heyd2003hybrid}.\ For band structure calculations, energy cut-off of 400 eV and a k-mesh of 11$\times$11$\times$7 are used.\ The system is optimized with PBE functionals and relaxed until a minimum force of 0.001 eV.\AA$^{-1}$ is reached for each atom in the unit cell.\ Phonon-dispersion spectrum is calculated using Phonopy package to confirm the dynamical stability of the proposed heterostructure \cite{togo2015first}.\ The optimized crystal structure is used to carry out the calculations of band structure with and without spin-orbit coupling (SOC).\ To confirm the non-trivial topological phase, Z$_{2}$ topological invariant is calculated using Kane and Mele formula  \cite{PhysRevLett.98.106803,PhysRevB.76.045302}.\ Further, maximally-localized Wannier functions (MLWF) are used to construct a tight-binding model using Wannier90 package \cite{RevModPhys.84.1419,mostofi2014updated}, which is used to calculate the Fermi surface for determining the the ratio of \textit{n$_{e}$/n$_{h}$}.\ The surface dispersion spectrum is calculated using Green's function iterative approach as implemented in Wannier tools \cite{PhysRevB.23.4997,sancho1985highly,WU2018405}.

\section{Results and discussions}
\begin{figure}[H]
\centering
 \includegraphics{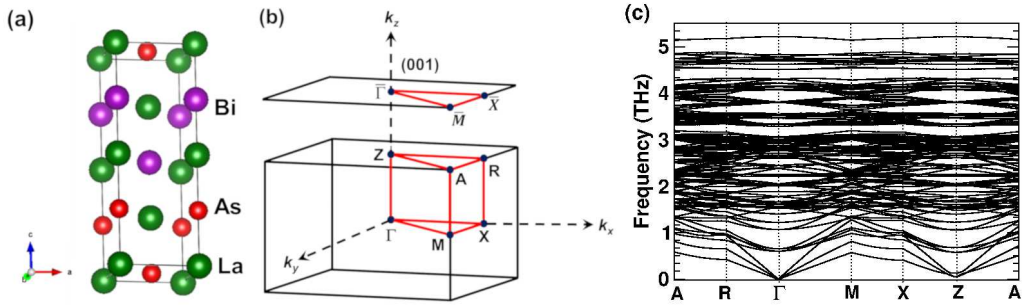}
  \caption{(color online) (a) 1$\times$1$\times$2 supercell of the proposed heterostructure of LaAs and LaBi.\ Green, red, and violet spheres denote `La', `As', and `Bi' atoms, respectively.\ (b) First Brillouin zone of simple tetragonal structure and its projection on (001) surface.\ (c) Phonon dispersion spectrum of LaAs/LaBi heterostructure.}
  \label{fgr:str1}
\end{figure}

To design the heterostructure, we consider alternate stacking of LaAs and LaBi unit cells along [001] direction and formed its 1$\times$1$\times$2 supercell as shown in Figure \ref{fgr:str1}(a).\ The heterostructure is found to exhibit a centrosymmetric tetragonal crystal structure having a space group of 129 (P4/nmm).\ The lattice parameters after optimizing the crystal structure are found to be `\textit{a}' = 4.559 \AA \ and `\textit{c}' = 12.825 \AA.\ In order to confirm the dynamical stability, we performed its phonon dispersion calculations [Figure \ref{fgr:str1}(c)] by constructing 2$\times$2$\times$2 supercell for the equilibrium lattice parameters.\ No imaginary phonon frequency is found in the phononic spectrum, indicating that the system is dynamically stable.\\

\subsection{\label{sec:level2}Band structure and Z$_{2}$ topological invariant}
\begin{figure}[htb!]
\centering
 \includegraphics{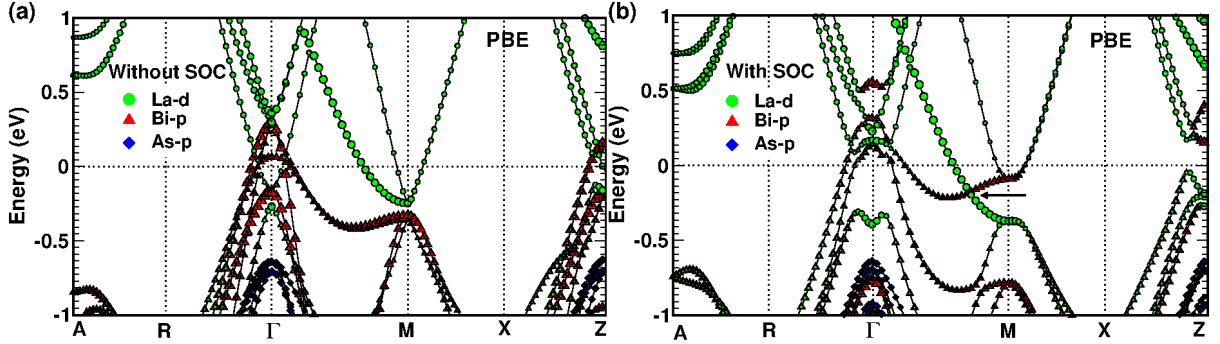}
  \caption{(color online) Bulk band structure of LaAs/LaBi heterostructure calculated using PBE functionals (a) without SOC and (b) with SOC.\ Green spheres, red triangles, and blue diamonds show the contribution of La-\textit{d}, Bi-\textit{p}, and As-\textit{p} orbitals, respectively.\ Arrow indicates the point of band crossing.}
  \label{fgr:str2}
\end{figure}

To determine the topological properties of the proposed heterostructure, first we have plotted its band structure using PBE functionals with and without including the SOC effect (Figure \ref{fgr:str2}).\ It can be seen that valence band and conduction band overlap with each other in both with and without SOC calculations, and are found to be semi-metallic.\ In addition, band structure plots suggest that near the Fermi level only Bi-\textit{p} orbitals (shown by red triangles) and La-\textit{d} orbitals (shown by green spheres) contribute in the valence band and conduction band, respectively; while the contribution of As-\textit{p} orbitals (shown by blue diamonds) lie well below the Fermi level.\ Moreover, a finite gap of 0.08 eV between valence band maxima and conduction band minima is observed at M point in band structure calculated without SOC.\ However on the inclusion of SOC, a band crossing between valence band and conduction band takes place along $\Gamma$-M direction, dictating to a band inversion between La-\textit{d} and Bi-\textit{p} orbitals.\

\begin{figure}
\centering
 \includegraphics{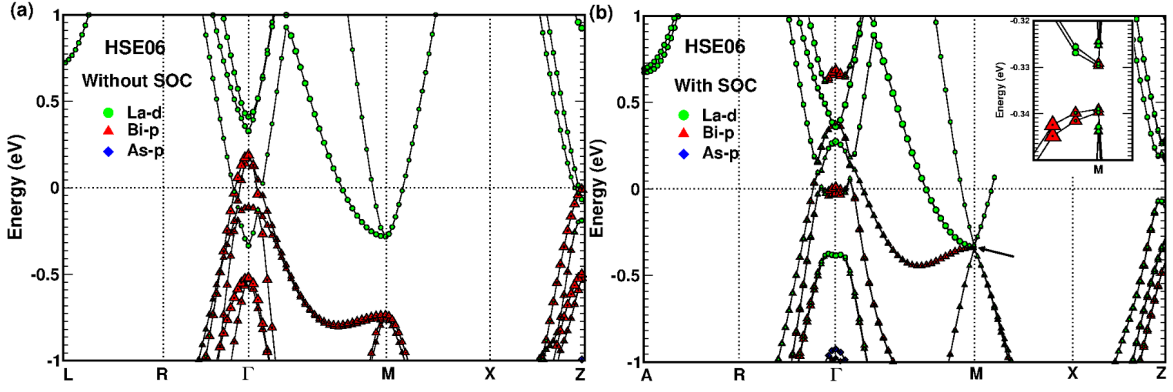}
  \caption{(color online) Bulk band structure of LaAs/LaBi heterostructure calculated using HSE06 functionals (a) without SOC and (b) with SOC.\ Green spheres, red triangles, and blue diamonds show the contribution of La-\textit{d}, Bi-\textit{p}, and As-\textit{p} orbitals, respectively.\ Inset shows the enlarged view of bands near the region indicated by an arrow.}
  \label{fgr:str3}
\end{figure}

Since GGA underestimates the gap and may overestimate the band inversion, therefore we have plotted the band structure computed using HSE06 functionals with and without SOC as depicted in Figure \ref{fgr:str3}.\ From band structure calculated without SOC [Figure \ref{fgr:str3} (a)], a finite gap of 0.451 eV is observed between valence band maxima and conduction band minima at M point, which is much greater than the one calculated using PBE functionals without SOC.\ Therefore, on the inclusion of SOC, the overlap between valence band and conduction band near M point is much less for HSE06 functionals than the PBE functionals and valence band maxima and conduction band minima near the Fermi level touch each other at M point [Figure \ref{fgr:str3} (b)].\ On close investigation, a small exchange between La-\textit{d} and Bi-\textit{p} orbitals is observed, indicating a band inversion at M point.\ It signifies that LaAs/LaBi heterostructure is topologically non-trivial.\ Since, only La-\textit{d} and Bi-\textit{p} orbitals are participating in band inversion for both PBE and HSE06 band structures with SOC [Figure \ref{fgr:str2} (b), Figure \ref{fgr:str3} (b)], therefore it can be interpreted that LaBi is playing prominent role in the non-trivial topological behaviour of LaAs/LaBi heterostructure.\\

Further, the presence of space inversion as well as time-reversal symmetry in the designed heterostructure enables us to calculate the Z$_{2}$ topological invariant as suggested by Kane and Mele \cite{PhysRevLett.98.106803,PhysRevB.76.045302}.\ A non-trivial topological phase in 3D is characterized by four Z$_{2}$ indices, i.e., (${\nu_{0}}$;${\nu_{1}}$${\nu_{2}}$${\nu_{3}}$).\ The value of first Z$_{2}$ index (${\nu_{0}}$) distinguishes the two distinct classes of strong and weak topological phases and can be calculated from the equation
\begin{equation}
(-1)^{\nu_{0}}=\prod_{i}\delta _{i}
\end{equation}

where $\prod_{i}\delta _{i}$ represents the parity product of all the occupied states at all time-reversal invariant momenta (TRIM) points in three-dimensional BZ.\ Table \ref{one} depicts the parity product at all eight TRIM points in the BZ of LaAs/LaBi heterostructure for PBE and HSE06 functionals with SOC.\\

\begin{table}[htb!]
\centering
\caption{Parity product at all TRIM points in the BZ for LaAs/LaBi heterostructure using PBE and HSE06 functionals with spin-orbit coupling.}
\label{one}
\begin{tabular}{|c|c|c|c|}
\hline
\textbf{Functional} & \textbf{TRIM points} & \textbf{$\delta_{m}$} & \textbf{$\nu_{0}$}  \\ \hline
\multirow{6}{*}{PBE+SOC} & 1A & + & \multirow{6}{*}{\textbf{1}} \\ \cline{2-3}
 & 2R & + &  \\ \cline{2-3}
 & 1$\Gamma$ & - &  \\ \cline{2-3}
 & 1M & - &  \\ \cline{2-3}
 & 2X & + &  \\ \cline{2-3}
 & 1Z & - &  \\ \hline
\multirow{6}{*}{HSE06+SOC} & 1A & + & \multirow{6}{*}{\textbf{1}} \\ \cline{2-3}
 & 2R & + &  \\ \cline{2-3}
 & 1$\Gamma$ & - &  \\ \cline{2-3}
 & 1M & - &  \\ \cline{2-3}
 & 2X & + &  \\ \cline{2-3}
 & 1Z & - &  \\ \hline
\end{tabular}
\end{table}

 It is observed that the value of ${\nu_{0}}$ for the heterostructure using PBE as well as HSE06 including SOC turns out to be 1, indicating that the system is topologically non-trivial.\\
 
It is to be noted that here we considered a heterostructure made up of two layers (N = 2) of LaAs and LaBi.\ However, on increasing the number of layers, we found that heterostructures made up of odd number of layers possess a centrosymmetric tetragonal crystal structure having a space group of 123 (P4/mmm), while heterostructures made up of even number of layers possess a centrosymmetric tetragonal crystal structure with a space group of 129 (P4/nmm).\ Meanwhile, the topological behaviour of their heterostructures for both even and odd layers remains non-trivial.\ The band structures for the heterostructures formed by N = 3 and 4 layers of the corresponding LaAs and LaBi are provided in the Supplementary Material 1.\

\subsection{\label{sec:level2}Surface state spectrum}
\begin{figure}[htb!]
\centering
 \includegraphics[width = 11cm]{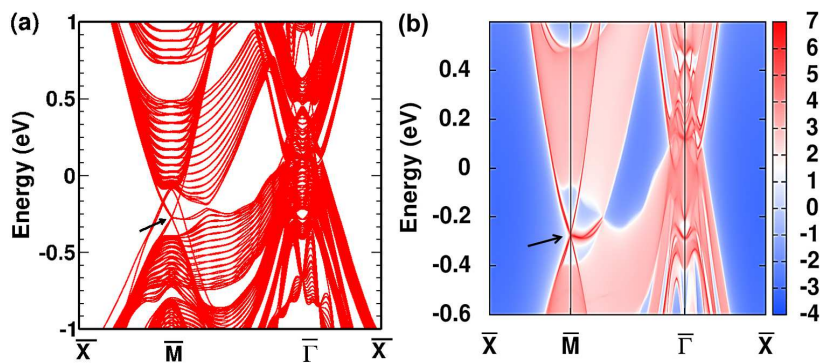}
  \caption{(color online) (a) Surface band structure and (b) momentum-resolved projected density of states (bottom surface) along [001] direction of LaAs/LaBi heterostructure calculated using PBE functionals with spin-orbit coupling.\ A Dirac point is indicated by an arrow.}
  \label{fgr:str4}
\end{figure}

To further investigate the non-trivial topological feature, we calculated the surface properties of the heterostructure by projecting its bulk on (001) surface using the slab method \cite{PhysRevB.23.4997,sancho1985highly,PhysRevB.23.4988}.\ Since, bulk band inversion is observed using both PBE and HSE06 functionals with SOC, and protected surface states are directly ensured by the presence of bulk band inversion, therefore we have used computationally less expensive PBE functionals to calculate the surface state spectrum.\ From our DFT calculations, it is found that only La-\textit{d} and Bi-\textit{p} orbitals are participating in the bulk band inversion [Figure \ref{fgr:str2} (b)], therefore we constructed the MLWFs only for La-\textit{d} and Bi-\textit{p} orbitals, and built the effective model Hamiltonian to calculate the surface state spectrum.\ The surface band structure and momentum-resolved projected density of states of a 15-layer slab along [001] direction of LaAs/LaBi heterostructure are plotted in Figure \ref{fgr:str4}.\

In the bulk spectrum, a crossing point is observed along $\Gamma$-M direction [Figure \ref{fgr:str2} (b)], while on projecting the bulk bands on (001) surface, a solid Dirac cone is observed at $\bar{M}$ point (Figure \ref{fgr:str4}), indicating that the surface of LaAs/LaBi heterostructure is topologically protected.\

\subsection{\label{sec:level2}Calculation of \textit{n$_{e}$/n$_{h}$} ratio}
\begin{figure}[H]
\centering
 \includegraphics{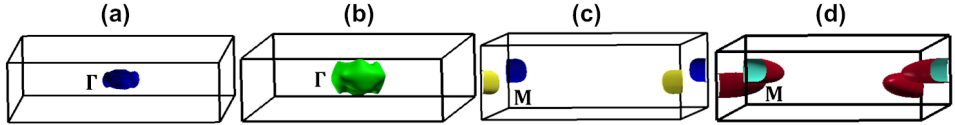}
  \caption{(color online) Fermi surface of LaAs/LaBi heterostructure.\ (a) and (b) represent the hole pockets at $\Gamma$ point, while (c) and (d) represent the electron pockets at M point.}
  \label{fgr:str5}
\end{figure}

In order to determine the \textit{n$_{e}$/n$_{h}$} ratio of the heterostucture, we plotted its Fermi surface using a dense k-grid of 150$\times$150$\times$50 (Figure \ref{fgr:str5}).\ From the band structure plot calculated using PBE functionals with SOC [Figure \ref{fgr:str2} (b)], we found that two of the valence bands cross the Fermi level at $\Gamma$ point, thereby leading to the two hole pockets at $\Gamma$ point.\ Similarly, at M point two of the conduction bands cross the Fermi level, therefore forming two electron pockets at M point (Figure \ref{fgr:str5}).\ The value of electron and hole densities calculated by finding the volume of their corresponding pockets are 2.598 $\times$ 10$^{20}$ cm$^{-3}$ and 2.40 $\times$ 10$^{20}$ cm$^{-3}$, respectively, and their ratio (\textit{n$_{e}$/n$_{h}$}) turns out to be 1.082.\ Further, it is also observed that broadening of the k-mesh for the construction of MLWFs leads to more accurate $n_{e}/n_{h}$ values, but requires a huge computational cost (details are provided in Supplementary Material 2).\\ 

Since LaAs and LaBi show XMR effect with perfect electron-hole compensation \cite{sun2016large,PhysRevB.96.235128,PhysRevB.93.235142}, and their heterostructure is also showing nearly perfect electron-hole compensation, therefore LaAs/LaBi heterostructure is also a possible candidate for XMR studies.\\

\section{Summary}
It is summarized that LaAs/LaBi heterostructure is topologically non-trivial with a surface Dirac cone along [001] direction.\ In addition, its electron and hole pockets are also found to be nearly compensated, facilitating it to be a promising candidate for exhibiting XMR effect.\ Therefore, the study provides an ideal candidate for a new topological heterostructure with XMR effect, which may have promising applications in realizing novel topological devices.

\section*{Acknowledgement} 
We acknowledge IIT Ropar for providing computational facility.

\section*{References}
\bibliographystyle{iopart-num}
\bibliography{Reference}

\providecommand{\newblock}{}
\begin{thebibliography}{10}
\expandafter\ifx\csname url\endcsname\relax
  \def\url#1{{\tt #1}}\fi
\expandafter\ifx\csname urlprefix\endcsname\relax\def\urlprefix{URL }\fi
\providecommand{\eprint}[2][]{\url{#2}}

\bibitem{RevModPhys.82.3045}
Hasan M~Z and Kane C~L 2010 {\em Rev. Mod. Phys.\/} {\bf 82}(4) 3045--3067
  \urlprefix\url{https://link.aps.org/doi/10.1103/RevModPhys.82.3045}

\bibitem{RevModPhys.83.1057}
Qi X~L and Zhang S~C 2011 {\em Rev. Mod. Phys.\/} {\bf 83}(4) 1057--1110
  \urlprefix\url{https://link.aps.org/doi/10.1103/RevModPhys.83.1057}

\bibitem{PhysRevLett.118.236402}
Agarwala A and Shenoy V~B 2017 {\em Phys. Rev. Lett.\/} {\bf 118}(23) 236402
  \urlprefix\url{https://link.aps.org/doi/10.1103/PhysRevLett.118.236402}

\bibitem{agarwala2019higher}
Agarwala A, Juri\ifmmode \check{c}\else \v{c}\fi{}i\ifmmode~\acute{c}\else
  \'{c}\fi{} V and Roy B 2020 {\em Phys. Rev. Research (R)\/} {\bf 2}(1) 012067
  \urlprefix\url{https://link.aps.org/doi/10.1103/PhysRevResearch.2.012067}

\bibitem{PhysRevLett.106.106802}
Fu L 2011 {\em Phys. Rev. Lett.\/} {\bf 106}(10) 106802
  \urlprefix\url{https://link.aps.org/doi/10.1103/PhysRevLett.106.106802}

\bibitem{armitage2018weyl}
Armitage N~P, Mele E~J and Vishwanath A 2018 {\em Rev. Mod. Phys.\/} {\bf 90}
  015001

\bibitem{barik2018multiple}
Barik R~K, Shinde R and Singh A~K 2018 {\em J. Phys.: Condens. Matter\/} {\bf
  30} 375702

\bibitem{wadhwa2019first}
Wadhwa P, Kumar S, Shukla A and Kumar R 2019 {\em J. Phys.: Condens. Matter\/}
  {\bf 31} 335401

\bibitem{PhysRevB.99.205112}
Mondal C, Barman C~K, Alam A and Pathak B 2019 {\em Phys. Rev. B\/} {\bf
  99}(20) 205112
  \urlprefix\url{https://link.aps.org/doi/10.1103/PhysRevB.99.205112}

\bibitem{PhysRevLett.121.086804}
Politano A, Chiarello G, Ghosh B, Sadhukhan K, Kuo C~N, Lue C~S, Pellegrini V
  and Agarwal A 2018 {\em Phys. Rev. Lett.\/} {\bf 121}(8) 086804
  \urlprefix\url{https://link.aps.org/doi/10.1103/PhysRevLett.121.086804}

\bibitem{PhysRevLett.121.226401}
Singh B, Ghosh B, Su C, Lin H, Agarwal A and Bansil A 2018 {\em Phys. Rev.
  Lett.\/} {\bf 121}(22) 226401
  \urlprefix\url{https://link.aps.org/doi/10.1103/PhysRevLett.121.226401}

\bibitem{PhysRevB.95.155420}
Huang G~Y and Xu H~Q 2017 {\em Phys. Rev. B\/} {\bf 95}(15) 155420
  \urlprefix\url{https://link.aps.org/doi/10.1103/PhysRevB.95.155420}

\bibitem{li2016negative}
Li H, He H, Lu H~Z, Zhang H, Liu H, Ma R, Fan Z, Shen S~Q and Wang J 2016 {\em
  Nat. Commun.\/} {\bf 7} 10301

\bibitem{PhysRevX.5.031023}
Huang X, Zhao L, Long Y, Wang P, Chen D, Yang Z, Liang H, Xue M, Weng H, Fang
  Z, Dai X and Chen G 2015 {\em Phys. Rev. X\/} {\bf 5}(3) 031023
  \urlprefix\url{https://link.aps.org/doi/10.1103/PhysRevX.5.031023}

\bibitem{vazifeh2013electromagnetic}
Vazifeh M~M and Franz M 2013 {\em Phys. Rev. Lett.\/} {\bf 111} 027201

\bibitem{zhang2011band}
Zhang J, Chang C~Z, Zhang Z, Wen J, Feng X, Li K, Liu M, He K, Wang L, Chen X
  {\em et~al.\/} 2011 {\em Nat. Commun.\/} {\bf 2} 574

\bibitem{kong2011ambipolar}
Kong D, Chen Y, Cha J~J, Zhang Q, Analytis J~G, Lai K, Liu Z, Hong S~S, Koski
  K~J, Mo S~K {\em et~al.\/} 2011 {\em Nat. Nanotechnol.\/} {\bf 6} 705

\bibitem{PhysRevLett.115.136801}
Chang C~Z, Tang P, Feng X, Li K, Ma X~C, Duan W, He K and Xue Q~K 2015 {\em
  Phys. Rev. Lett.\/} {\bf 115}(13) 136801
  \urlprefix\url{https://link.aps.org/doi/10.1103/PhysRevLett.115.136801}

\bibitem{eschbach2015realization}
Eschbach M, M{\l}y{\'n}czak E, Kellner J, Kampmeier J, Lanius M, Neumann E,
  Weyrich C, Gehlmann M, Gospodari{\v{c}} P, D{\"o}ring S {\em et~al.\/} 2015
  {\em Nat. Commun.\/} {\bf 6} 8816

\bibitem{dey2018bulk}
Dey U, Chakraborty M, Taraphder A and Tewari S 2018 {\em Sci. Rep.\/} {\bf 8}
  14867

\bibitem{ali2014large}
Ali M~N, Xiong J, Flynn S, Tao J, Gibson Q~D, Schoop L~M, Liang T,
  Haldolaarachchige N, Hirschberger M, Ong N {\em et~al.\/} 2014 {\em Nature\/}
  {\bf 514} 205

\bibitem{jiang2015signature}
Jiang J, Tang F, Pan X~C, Liu H~M, Niu X~H, Wang Y~X, Xu D~F, Yang H~F, Xie
  B~P, Song F~Q, Dudin P, Kim T~K, Hoesch M, Das P~K, Vobornik I, Wan X~G and
  Feng D~L 2015 {\em Phys. Rev. Lett.\/} {\bf 115} 166601

\bibitem{PhysRevB.94.041103}
Wang Y~Y, Yu Q~H, Guo P~J, Liu K and Xia T~L 2016 {\em Phys. Rev. B\/} {\bf
  94}(4) 041103(R)
  \urlprefix\url{https://link.aps.org/doi/10.1103/PhysRevB.94.041103}

\bibitem{neupane2016observation}
Neupane M, Hosen M~M, Belopolski I, Wakeham N, Dimitri K, Dhakal N, Zhu J~X,
  Hasan M~Z, Bauer E~D and Ronning F 2016 {\em J. Phys.: Condens. Matter\/}
  {\bf 28} 23LT02

\bibitem{ghimire2016magnetotransport}
Ghimire N, Botana A, Phelan D, Zheng H and Mitchell J~F 2016 {\em J. Phys.:
  Condens. Matter\/} {\bf 28} 235601

\bibitem{PhysRevB.95.115140}
Lou R, Fu B~B, Xu Q~N, Guo P~J, Kong L~Y, Zeng L~K, Ma J~Z, Richard P, Fang C,
  Huang Y~B, Sun S~S, Wang Q, Wang L, Shi Y~G, Lei H~C, Liu K, Weng H~M, Qian
  T, Ding H and Wang S~C 2017 {\em Phys. Rev. B\/} {\bf 95}(11) 115140
  \urlprefix\url{https://link.aps.org/doi/10.1103/PhysRevB.95.115140}

\bibitem{yu2017magnetoresistance}
Yu Q~H, Wang Y~Y, Lou R, Guo P~J, Xu S, Liu K, Wang S and Xia T~L 2017 {\em
  Europhys. Lett.\/} {\bf 119} 17002

\bibitem{wang2018topological}
Wang Y, Liang D, Ge M, Yang J, Gong J, Luo L, Pi L, Zhu W, Zhang C and Zhang Y
  2018 {\em J. Phys.: Condens. Matter\/} {\bf 30} 155701

\bibitem{PhysRevB.99.245131}
Vashist A, Gopal R~K, Srivastava D, Karppinen M and Singh Y 2019 {\em Phys.
  Rev. B\/} {\bf 99}(24) 245131
  \urlprefix\url{https://link.aps.org/doi/10.1103/PhysRevB.99.245131}

\bibitem{liang2015ultrahigh}
Liang T, Gibson Q, Ali M~N, Liu M, Cava R~J and Ong N~P 2015 {\em Nat.
  Mater.\/} {\bf 14} 280

\bibitem{PhysRevLett.113.216601}
Pletikosi\ifmmode~\acute{c}\else \'{c}\fi{} I, Ali M~N, Fedorov A~V, Cava R~J
  and Valla T 2014 {\em Phys. Rev. Lett.\/} {\bf 113}(21) 216601
  \urlprefix\url{https://link.aps.org/doi/10.1103/PhysRevLett.113.216601}

\bibitem{sun2016large}
Sun S, Wang Q, Guo P~J, Liu K and Lei H 2016 {\em New J. Phys.\/} {\bf 18}
  082002

\bibitem{PhysRevB.96.235128}
Yang H~Y, Nummy T, Li H, Jaszewski S, Abramchuk M, Dessau D~S and Tafti F 2017
  {\em Phys. Rev. B\/} {\bf 96}(23) 235128
  \urlprefix\url{https://link.aps.org/doi/10.1103/PhysRevB.96.235128}

\bibitem{PhysRevB.93.235142}
Guo P~J, Yang H~C, Zhang B~J, Liu K and Lu Z~Y 2016 {\em Phys. Rev. B\/} {\bf
  93}(23) 235142
  \urlprefix\url{https://link.aps.org/doi/10.1103/PhysRevB.93.235142}

\bibitem{PhysRevB.98.220102}
Khalid S, Sabino F~P and Janotti A 2018 {\em Phys. Rev. B\/} {\bf 98}(22)
  220102(R) \urlprefix\url{https://link.aps.org/doi/10.1103/PhysRevB.98.220102}

\bibitem{duan2018tunable}
Duan X, Wu F, Chen J, Zhang P, Liu Y, Yuan H and Cao C 2018 {\em Commun.
  Phys.\/} {\bf 1} 71

\bibitem{PhysRevB.54.11169}
Kresse G and Furthm\"uller J 1996 {\em Phys. Rev. B\/} {\bf 54}(16)
  11169--11186
  \urlprefix\url{https://link.aps.org/doi/10.1103/PhysRevB.54.11169}

\bibitem{kresse1999ultrasoft}
Kresse G and Joubert D 1999 {\em Phys. Rev. B\/} {\bf 59} 1758

\bibitem{PhysRevLett.77.3865}
Perdew J~P, Burke K and Ernzerhof M 1996 {\em Phys. Rev. Lett.\/} {\bf 77}(18)
  3865--3868
  \urlprefix\url{https://link.aps.org/doi/10.1103/PhysRevLett.77.3865}

\bibitem{heyd2003hybrid}
Heyd J, Scuseria G~E and Ernzerhof M 2003 {\em J. Chem. Phys.\/} {\bf 118}
  8207--8215

\bibitem{togo2015first}
Togo A and Tanaka I 2015 {\em Scr. Mater.\/} {\bf 108} 1--5

\bibitem{PhysRevLett.98.106803}
Fu L, Kane C~L and Mele E~J 2007 {\em Phys. Rev. Lett.\/} {\bf 98}(10) 106803
  \urlprefix\url{https://link.aps.org/doi/10.1103/PhysRevLett.98.106803}

\bibitem{PhysRevB.76.045302}
Fu L and Kane C~L 2007 {\em Phys. Rev. B\/} {\bf 76}(4) 045302
  \urlprefix\url{https://link.aps.org/doi/10.1103/PhysRevB.76.045302}

\bibitem{RevModPhys.84.1419}
Marzari N, Mostofi A~A, Yates J~R, Souza I and Vanderbilt D 2012 {\em Rev. Mod.
  Phys.\/} {\bf 84}(4) 1419--1475
  \urlprefix\url{https://link.aps.org/doi/10.1103/RevModPhys.84.1419}

\bibitem{mostofi2014updated}
Mostofi A~A, Yates J~R, Pizzi G, Lee Y~S, Souza I, Vanderbilt D and Marzari N
  2014 {\em Comput. Phys. Commun.\/} {\bf 185} 2309--2310

\bibitem{PhysRevB.23.4997}
Lee D~H and Joannopoulos J~D 1981 {\em Phys. Rev. B\/} {\bf 23}(10) 4997--5004
  \urlprefix\url{https://link.aps.org/doi/10.1103/PhysRevB.23.4997}

\bibitem{sancho1985highly}
Sancho M~L, Sancho J~L, Sancho J~L and Rubio J 1985 {\em J. Phys. F: Met.
  Phys.\/} {\bf 15} 851

\bibitem{WU2018405}
Wu Q, Zhang S, Song H~F, Troyer M and Soluyanov A~A 2018 {\em Comput. Phys.
  Commun.\/} {\bf 224} 405 -- 416 ISSN 0010-4655
  \urlprefix\url{http://www.sciencedirect.com/science/article/pii/S0010465517303442}

\bibitem{PhysRevB.23.4988}
Lee D~H and Joannopoulos J~D 1981 {\em Phys. Rev. B\/} {\bf 23}(10) 4988--4996
  \urlprefix\url{https://link.aps.org/doi/10.1103/PhysRevB.23.4988}

\end{thebibliography}



\end{document}